\documentclass[preprint,12pt]{elsarticle}

\usepackage{amsmath}

\usepackage{amssymb}

\newcommand{\dps}{\displaystyle}

\newcommand{\be}{\begin{equation}} 
\newcommand{\ee}{\end{equation}} 
\newcommand{\bea}{\begin{eqnarray}} 
\newcommand{\eea}{\end{eqnarray}}

\def\slash#1{#1\!\!\!\raise.15ex\hbox {/}}
\newcommand{\slD}{\,\raise.15ex\hbox{$/$}\kern-.27em\hbox{$\!\!\!D$}}
\newcommand{\slpartial}{\raise.15ex\hbox{$/$}\kern-.57em\hbox{$\partial$}}

\def\2int{\int_{0}^{1}  \!\!\! du \! \int_{0}^{1}  \!\!\! dv}

\def\e{\mbox{e}}

\def\Eins{{\mathchoice {\rm 1\mskip-4mu l} {\rm 1\mskip-4mu l}
{\rm 1\mskip-4.5mu l} {\rm 1\mskip-5mu l}}}
\def\Z{{\mathchoice {\hbox{$\sf\textstyle Z\kern-0.4em Z$}}
{\hbox{$\sf\textstyle Z\kern-0.4em Z$}}
{\hbox{$\sf\scriptstyle Z\kern-0.3em Z$}}
{\hbox{$\sf\scriptscriptstyle Z\kern-0.2em Z$}}}}

\def\no{\noindent}
\def\non{\nonumber\\}

\def\Eins{{\mathchoice {\rm 1\mskip-4mu l} {\rm 1\mskip-4mu l}{\rm 1\mskip-4.5mu l} {\rm 1\mskip-5mu l}}}

\def\half{\frac{1}{2}}

\def\mn{{\mu\nu}}

\def\e{\,{\rm e}}

\def\b0{{\bf 0}}

\def\bear{\begin{eqnarray}}
\def\ear{\end{eqnarray}\noindent}
\def\bec{\blue\begin{equation}}
\def\eec{\end{equation}\black\noindent}
\def\bearc{\blue\begin{eqnarray}}
\def\earc{\end{eqnarray}\black\noindent}
\def\benn{\begin{enumerate}}
\def\enn{\end{enumerate}}

\def\totint{\int_{-\infty}^{\infty}}

\def\intdp3{\int\frac{d^3p}{(2\pi)^3}}
\def\intdp4{\int\frac{d^4p}{(2\pi)^4}}

\def\kinb{\frac{\dot x^2}{4}}

\def\4piTD{{(4\pi T)}^{-{D\over 2}}}
\def\4piT4{{(4\pi T)}^{-2}}

\newcommand{\slG}{{{\dot G}\!\!\!\! \raise.15ex\hbox {/}}}

\def\ddtau{{d\over d\tau}}

\def\dps{\displaystyle}

\def\Nint{\prod_{i=1}^N\int_0^Td\tau_i}

\begin{document}

\begin{frontmatter}



\title{
Master formulas for photon amplitudes in a combined constant and plane-wave background field
}

\author[label1]{Christian Schubert}

\affiliation[label1]{organization={Extreme Light Infrastructure ERIC},
          addressline={Za Radnici 835},
           city={Dolni Brezany},
                      postcode={25241},
          country={Czech Republic\\
        Christian.Schubert@eli-beams.eu, Rashid.Shaisultanov@eli-beams.eu}}


\author[label1]{Rashid Shaisultanov}


\begin{abstract}
The worldline formalism has previously been used for deriving compact master formulas for the QED $N$ - photon amplitudes 
in vacuum, in a constant field and in a plane-wave field. Here we carry this program one step further by deriving master formulas
for the scalar and spinor QED $N$-photon amplitudes in the background of the ``parallel'' special case of a combined 
constant and plane-wave field.
\end{abstract}
%
%
%
\begin{keyword}
worldline formalism \sep photon amplitudes \sep plane-wave background \sep constant field



\end{keyword}

\end{frontmatter}


\section{Introduction}
\label{intro}

The analytical performance of calculations in strong-field QED requires one to know the exact Dirac propagator in the field,
which is available only for a few field configurations. Among those, the most important ones are the constant field and the
plane-wave one (see, e.g., \cite{ritus-ginzburg,frgish-book,ditgie-book}). 
Nevertheless, since the exact propagators have a very complicated structure such calculations tend to be extremely lengthy and tedious,
and despite of decades of effort the technology of Feynman diagrams with external fields is presently still stuck at the level
of one-loop amplitudes with not
more than three fermion propagators
\cite{volkov,furry,schwinger51,nikrit1,nikrit2,baibre,adler71,tsai,bks1,bks2,bkms,bks3,bekmit,monari,akhmedov,bamish,bamishPRL,dimike,dipmil,dipiazza,dhimt,mekedi,mimefe,poddip,pialop}. 

Clearly one would prefer to have a method that would avoid the segmentation of fermion loops and lines into
individual propagators. Such an approach, based on relativistic worldline path integrals, has been developed by Feynman  \cite{feynman:pr80,feynman:pr84}
in the early fifties, but gained traction only around 1990 following developments in string theory \cite{berkosNPB,strassler1} (for reviews, see \cite{41,Edwards:2019eby}). 
Although equivalent to the standard Feynman diagram approach, methodically it is quite different since each of the path integrals represents a whole fermion loop or line,
rather than individual propagators, and the emission or absorption of a photon by the fermion is effected through the insertion of a vertex operator, as in string theory.  
As far as QED is concerned, the central formula of the formalism is the following master formula for the one-loop $N$-photon amplitude in scalar QED
\cite{polyakov-book,berkosNPB,strassler1}:
\begin{eqnarray}
\Gamma_{\rm scal}(k_1,\varepsilon_1;\ldots;k_N,\varepsilon_N)
&=&
{(-ie)}^N
{(2\pi )}^D\delta (\sum k_i)
{\dps\int_{0}^{\infty}}\frac{dT}{T}
{(4\pi T)}^{-\frac{D}{2}}
e^{-m^2T}
\nonumber\\
&& \hspace{-150pt}
\times
\prod_{i=1}^N \int_0^T 
d\tau_i
\exp\biggl\lbrace\sum_{i,j=1}^N 
\Bigl\lbrack  \half G_{ij} k_i\cdot k_j
-i\dot G_{ij}\varepsilon_i\cdot k_j
+\half\ddot G_{ij}\varepsilon_i\cdot\varepsilon_j
\Bigr\rbrack\biggr\rbrace
\Bigl\vert_{\varepsilon_1\varepsilon_2\ldots \varepsilon_N}
\, .
\label{scalarqedmaster}
\end{eqnarray}
\no
Here we have abbreviated $G_{ij}\equiv G(\tau_i,\tau_j)$ etc., where $G$ is the ``bosonic worldline Green's function'' defined by
\bear
G(\tau,\tau') \equiv \mid \tau-\tau'\mid 
-\frac{(\tau-\tau')^2}{T} 
\label{defG}
\ear
and a `dot' denotes a derivative acting on the first variable,
\begin{eqnarray}
\dot G(\tau,\tau') &=& {\rm sgn}(\tau - \tau')
- 2 \frac{(\tau - \tau')}{T}, \quad
\ddot G(\tau,\tau')
= 2 {\delta}(\tau - \tau')
- \frac{2}{T}
\, .
\label{GdGdd}
\end{eqnarray}
\noindent
The exponential must still be expanded and only the terms be retained that contain
each polarisation vector $\varepsilon_i$ linearly:
\bear 
\exp\bigl\lbrace 
\cdot
\bigr\rbrace 
\bigl\vert_{\varepsilon_1\varepsilon_2\ldots \varepsilon_N}
 \quad\equiv
 {(-i)}^N P_N(\dot G_{ij},\ddot G_{ij})
 \exp\biggl[\half \sum_{i,j=1}^N G_{ij}k_i\cdot
k_j \biggr] 
\label{defPN}
\ear
with certain polynomials $P_N$. 

The generalization to the spinor-loop case in the modern approach is done through the addition
of a Grassmann path integral, to be evaluated with the ``fermionic worldline Green's function'' $G_F(\tau,\tau') \equiv {\rm sgn}(\tau-\tau')$
\cite{fradkin,polyakov-book,berkosNPB,strassler1}.
Subsequently, it was discovered that this master formula can be generalized to the inclusion
of a constant external field in a very economical way, namely by a modification of the path-integral determinant \cite{5} and
the introduction of generalized worldline Green's functions ${\cal G}_B,{\cal G}_F$ that take the external field into account 
\cite{shaisultanov,18}
\begin{eqnarray}
{\cal G}_{B}(\tau,\tau') &\equiv& \frac{T}{2{\cal Z}^2}
\biggl({{\cal Z}\over{{\rm sin}{\cal Z}}}
\,{\rm e}^{-i{\cal Z}\dot G(\tau,\tau')}
+i{\cal Z}\dot G(\tau,\tau') -1\biggr)
\label{calGB}\\
{\cal G}_F(\tau,\tau') &\equiv&
G_F(\tau,\tau')
{{\rm e}^{-i{\cal Z}\dot G (\tau,\tau')}\over {\rm cos}{\cal Z}}
\label{calGF}
\end{eqnarray}
\noindent
where ${\cal Z}_{\mu\nu} \equiv eF_{\mu\nu}T$.
These expressions for the constant-field Green's functions
should be understood as power series in the Lorentz matrix $\cal Z$.
The master formula for the photon amplitudes in vacuum \eqref{scalarqedmaster} can then be generalized to the
constant field case as follows \cite{shaisultanov,18},
\begin{eqnarray}
&&\Gamma_{\rm scal}
[k_1,\varepsilon_1;\ldots;k_N,\varepsilon_N;F]
=
{(-ie)}^N
{(2\pi )}^D\delta^D (\sum k_i)
\label{masterF}
\\
&&\hspace{20pt}\times
{\dps\int_{0}^{\infty}}{dT\over T}
{(4\pi T)}^{-{D\over 2}}
e^{-m^2T}
{\rm det}^{{1\over 2}}
\biggl[
\frac{\cal Z}{{\rm sin}{\cal Z}}
\biggr]
\prod_{i=1}^N \int_0^T \!\!\!
d\tau_i
\nonumber\\
&&\hspace{20pt}\times
\exp\biggl\lbrace\sum_{i,j=1}^N 
\Bigl\lbrack \half k_i\cdot {\cal G}_{Bij}\cdot  k_j
-i\varepsilon_i\cdot\dot{\cal G}_{Bij}\cdot k_j
+\half
\varepsilon_i\cdot\ddot {\cal G}_{Bij}\cdot\varepsilon_j
\Bigr\rbrack\biggr\rbrace
\Big\vert_{\varepsilon_1\varepsilon_2\cdots \varepsilon_N}\,. 
\nonumber
\label{masterF}
\end{eqnarray}
\no
This representation has already been tested on the calculations of the tadpole \cite{112,113} and 
vacuum polarization amplitudes \cite{ditsha,40}, as well as on
magnetic photon splitting \cite{17,adler-book}, and found to be significantly more efficient than other available methods. 

More recently, the worldline approach has also been applied to the plane-wave case, defined by a vector potential $A(x)$ of the form
\bear
e A_{\mu}(x) = a_{\mu}(n\cdot x)
\ear
where $n^{\mu}$ is a null vector, $n^2=0$, and we will further impose the light-front gauge condition $n\cdot a = 0$.
This case  turned out to be less straightforward than the constant-field one, since the resulting path integrals are not 
manifestly gaussian. Nevertheless, building on work by A. Ilderton and G. Torgrimsson on the vacuum-polarization case \cite{ildtor},
it was shown by J.P. Edwards and one of the authors \cite{141} how to use the special kinematics of the plane-wave case to 
reduce the path integrals to gaussian ones, which then also made it straightforward to arrive at a master formula for the $N$-photon amplitudes
in such a background.  

The gaussian property is equivalent to the exactness of the semiclassical approximation for the constant-field and plane-wave backgrounds,
and it is natural to ask whether it continues to hold for the combination of both backgrounds. While this is not the case in general, there
is a special case where it applies, namely when there exists a Lorentz frame where the electric and magnetic fields are parallel, and
the plane wave propagates in the same direction. 
In such a field the Klein-Gordon and Dirac equations are still exactly solvable, as was first shown for the magnetic case by 
P.~J. Redmond \cite{redmond} in 1965 and for the general case by I.~A. Batalin and E.~S. Fradkin 
in 1970 \cite{batfra} (see also \cite{narnik,frgish-book}). This fact has prompted studies of the scalar \cite{baimil1} and spinor \cite{baimil2} mass operators in such a field, 
as well as of the spinor vacuum polarisation \cite{lobkha}.
Apart from its mathematical interest as the most general known field permitting the exact solution of the Klein-Gordon and Dirac equations, 
it leads also to a number of interesting physical effects related to the existence of a cyclotron resonance \cite{baimil1,baimil2} (see also \cite{kollev}).
 This resonance requires only the magnetic
field, and happens when the wave frequency coincides with the cyclotron frequency of a particle in the magnetic field (with the Doppler shift taken into account). 

In this letter, we use the worldline formalism to derive master formulas for the $N$-photon amplitudes in this field configuration, working in parallel on scalar and
spinor QED.
We start in the following section with a summary of the results of \cite{141} on the $N$-photon amplitudes in a plane-wave background. 

\section{$N$-photon amplitudes in a plane-wave background}
\label{pw}

The starting point of \cite{141} was the worldline path integral representation of the $N$-photon amplitude in scalar QED
in such a background \cite{feynman:pr80,polyakov-book,strassler1},
\bear
\Gamma_{\rm scal}(\lbrace{k_i,\varepsilon_i\rbrace};a)
 &=&
(-ie)^N 
\int_0^{\infty}
\frac{dT}{T}\,
\e^{-m^2T}
\int
Dx\,
\e^{-\int_0^T d\tau \bigl\lbrack\kinb +i \dot x^{\mu}a_{\mu}(n\cdot x)\bigr\rbrack}
\nonumber\\
&& \times
 V^{\gamma}_{\rm scal}[k_1,\varepsilon_1]\cdots V^{\gamma}_{\rm scal}[k_N,\varepsilon_N]
 \, .
\label{12-Nphotonpw}
\ear
Here the path integral runs over all closed trajectories in (euclidean) spacetime obeying the 
periodicity condition $x(T)=x(0)$ in proper time, and each photon is represented by the following {\sl photon vertex operator}:
\bear
V^{\gamma}_{\rm scal}[k,\varepsilon] &\equiv &\int_0^T d\tau \,\varepsilon\cdot\dot x(\tau)\,{\e}^{ik\cdot x(\tau)}
\, .
\label{defphotonvertop}
\ear\no
It is convenient to introduce (euclidean) light-cone coordinates \cite{ilderton}.
Setting $n^\mu \equiv \frac{1}{\sqrt{2}}(0,0,1,i)$ one denotes $x^+ \equiv n\cdot x = \frac{1}{\sqrt{2}}(x^3+ix^4)$ (``light-front time''), $x^- \equiv \frac{1}{\sqrt{2}}(- x^3+ix^4)$,
and $x^{\perp} \equiv (x^1,x^2)$. 

Before proceeding with the functional integrals one has to address
the existence of a zero mode, the constant paths, which is done by separating off the average position $x_0^\mu$ of the trajectory, $x^\mu(\tau)=x_0^\mu + q^\mu(\tau)$.
Defining also $k^\pm \equiv \frac{1}{\sqrt{2}} (\pm k^3 + ik^4)$,
and using the decomposition
\bear
k\cdot x = - k^+x^- - k^-x^+ + k^1x^1 + k^2x^2
\label{12-kx}
\ear
allows one to integrate out $x_0^\mu$ but for its $x_0^+$ component:
\bear
&&
\int
Dx\,
\e^{-\int_0^T d\tau \bigl\lbrack\kinb +i \dot x^{\mu}a_{\mu}(n\cdot x)\bigr\rbrack}
 V^{\gamma}_{\rm scal}[k_1,\varepsilon_1]\cdots V^{\gamma}_{\rm scal}[k_N,\varepsilon_N]
 \nonumber\\
 &&
 =
 (2\pi)^3
 \delta\bigl(\sum_{i=1}^N k_i^1\bigr)
\delta\bigl(\sum_{i=1}^N k_i^2\bigr)
\delta\bigl(\sum_{i=1}^N k_i^+\bigr)
\totint dx_0^+ 
\e^{-i x_0^+ \sum_{i=1}^N k_i^-}
\Nint
\nonumber\\&&\times
\int Dq 
\e^{-\int_0^T d\tau \bigl\lbrack\frac{\dot q^2}{4} +i \dot q^{\mu}a_{\mu}(x_0^+ + n\cdot q(\tau))\bigr\rbrack}
\,\e^{\sum_{i=1}^N (ik_i \cdot q_i + \varepsilon_i\cdot \dot q_i)}
\Big\vert_{\varepsilon_1\varepsilon_2\cdots \varepsilon_N}\,. 
\label{12-intx0}
\ear
The functional integral might seem intractable, since the integration variable $q^\mu (\tau)$ appears in the argument of the undetermined function $a_\mu$. 
The upshot is that the vanishing relations
\bear
n^2=n\cdot a=0
\label{idvanbasic}
\ear
can be used to replace the argument by a definite function of proper-time. This can be done in various ways \cite{ildtor,141}; here we will present a new approach
based on auxiliary functional integrals. Using the integral representation of the functional delta function, we can rewrite
\bear
\e^{-i\int_0^T d\tau \, \dot q^{\mu}a_{\mu}(x_0^+ + n\cdot q(\tau))}
&=&
\int D\phi (\tau) \prod_{\tau} \delta(\phi(\tau) - n\cdot q(\tau))
\e^{-i\int_0^T d\tau \, \dot q^{\mu}a_{\mu}(x_0^+ + \phi(\tau))}
\nonumber\\ &&\hspace{-70pt}
=
\int D\phi (\tau) 
\int D\chi (\tau) \
\e^{i\int_0^Td\tau \chi(\tau) \bigl(\phi(\tau)-n\cdot q(\tau)\bigr)}
\e^{-i\int_0^T d\tau \, \dot q^{\mu}a_{\mu}(x_0^+ + \phi(\tau))}
\, .
\nonumber\\
\ear
Employing this identity in \eqref{12-intx0}, the functional integral over $q^\mu (\tau)$ becomes gaussian,
and the usual ``completing-the-square'' procedure can be applied. This leads to
\bear
&&
\int Dq 
\e^{-\int_0^T d\tau \bigl\lbrack\frac{\dot q^2}{4} +i \dot q\cdot a(\tau) 
\bigr\rbrack}
\,\e^{\sum_{i=1}^N (ik_i \cdot q_i + \varepsilon_i\cdot \dot q_i)}
\e^{i\int_0^Td\tau \chi(\tau) \bigl(-n\cdot q(\tau)\bigr)}
\nonumber\\
&&
=
{(4\pi T)}^{-{D\over 2}}
\e^{-\frac{1}{2} \int_0^Td\tau\int_0^Td\tau' \ddot G(\tau,\tau')a(\tau)\cdot a(\tau')
+\sum_{i,j=1}^N \bigl\lbrack  \half G_{ij} k_i\cdot k_j
-i\dot G_{ij}\varepsilon_i\cdot k_j
+\half\ddot G_{ij}\varepsilon_i\cdot\varepsilon_j
\bigr\rbrack
}
\nonumber\\
&&\quad \times
\e^{
-\sum_{i=1}^N \int_0^Td\tau 
\bigl\lbrace a(\tau)\cdot
\bigl\lbrack \dot G(\tau,\tau_i) k_i + i \ddot G(\tau,\tau_i) \varepsilon_i \bigr\rbrack
+\chi(\tau) n\cdot 
\bigl\lbrack
G(\tau,\tau_i) k_i + i \dot G(\tau,\tau_i)\varepsilon_i
\bigr\rbrack
\bigr\rbrace
}
\label{12-qintfin}
\ear
where $a_\mu(\tau) \equiv a_{\mu}(x_0^+  +\phi(\tau))$ and the vanishing relations \eqref{idvanbasic}
have been used to eliminate some terms in the exponent, including one quadratic in $\chi(\tau)$. 
As a consequence, the functional integral over $\chi(\tau)$ becomes trivial and simply produces a 
$ \prod_{\tau} \delta\bigl(\phi(\tau) + i \sum_{i=1}^N n\cdot \bigl\lbrack G(\tau,\tau_i) k_i + i \dot G(\tau,\tau_i)\varepsilon_i\bigr\rbrack \bigr)$,
which then can be used to eliminate the integral over $\phi(\tau)$. 
The effect of all this is to change the argument of $a_\mu$ to
\bear
a_\mu(\tau)
=
a_\mu 
\Bigl(x_0^+  + n\cdot \sum_{i=1}^N[-i G(\tau,\tau_i)k_i + \dot G(\tau,\tau_i)\varepsilon_i]\Bigr)
\, .
\label{12-abuse}
\ear
Further, introducing the worldline average of a general function $f$
\bear
\langle\langle f \rangle\rangle
\equiv
\frac{1}{T} \int_0^Td\tau f(\tau)
\label{12-wa}
\ear
and using \eqref{GdGdd} one can rewrite
\bear
\half \int_0^Td\tau\int_0^Td\tau' \ddot G(\tau,\tau')a(\tau)\cdot a(\tau')
&=&
T\Bigl(\langle\langle a^2 \rangle\rangle - \langle\langle a \rangle\rangle^2\Bigr)
\, ,
\\
\sum_{i=1}^N \int_0^Td\tau \,
\ddot G(\tau,\tau_i) a(\tau)\cdot \varepsilon_i 
&=&
2 \sum_{i=1}^N\Bigl(a(\tau_i)-\langle\langle a\rangle\rangle \Bigr) \cdot \varepsilon_i
\, .
\ear
For the integral involving $\dot G(\tau,\tau_i)$, one can use integration-by-parts to rewrite
\bear
\sum_{i=1}^N \int_0^Td\tau \,\dot G(\tau,\tau_i) a(\tau)\cdot k_i
=
-2
\sum_{i=1}^N k_i \cdot 
\bigl(I(\tau_i)-\langle\langle I \rangle\rangle \bigr)
\label{IBPI}
\ear
with the periodic integral function
\bear
I_\mu(\tau) \equiv \int_0^\tau d\tau' \Bigl(a_\mu(\tau') - \langle\langle a_\mu\rangle\rangle \Bigr )
\, .
\ear
In this manner, one gets the following master formula for the scalar QED
$N$-photon amplitude in a plane-wave background \cite{141},
\bear
\Gamma_{\rm scal}(\lbrace{k_i,\varepsilon_i\rbrace};a) &=&
(-ie)^N 
(2\pi)^3 
\delta\bigl(\sum_{i=1}^N k_i^1\bigr)
\delta\bigl(\sum_{i=1}^N k_i^2\bigr)
\delta\bigl(\sum_{i=1}^N k_i^+\bigr)
\totint dx_0^+ \e^{-i x_0^+ \sum_{i=1}^N k_i^-}
\nonumber\\&&\hspace{-80pt}\times
\int_0^{\infty}
\frac{dT}{T}\,
{(4\pi T)}^{-{D\over 2}}
\prod_{i=1}^N \int_0^Td\tau_i
\e^{
\sum_{i,j=1}^N 
\bigl\lbrack  \half G_{ij} k_i\cdot k_j
-i\dot G_{ij}\varepsilon_i\cdot k_j
+\half\ddot G_{ij}\varepsilon_i\cdot\varepsilon_j
\bigr\rbrack}
\nonumber\\&&\hspace{-80pt}\times
\e^{-\bigl(m^2+ \langle\langle a^2 \rangle\rangle - \langle\langle a \rangle\rangle^2\bigr)T+2\sum_{i=1}^N k_i \cdot 
\bigl(I(\tau_i)-\langle\langle I \rangle\rangle \bigr)
-2i \sum_{i=1}^N\bigl(a(\tau_i)-\langle\langle a\rangle\rangle \bigr) \cdot \varepsilon_i
}
\Bigl\vert_{\varepsilon_1\cdots \varepsilon_N}
\, .
\nonumber\\
\label{Nphotonpw}
\ear
Further simplification can be achieved by choosing a gauge for the $\varepsilon_i$.
Imposing the condition
\bear
n\cdot \varepsilon_i = 0 \quad (i=1,\ldots,N) \, ,
\label{polgauge}
\ear
reduces \eqref{12-abuse} to
\bear
a_\mu(\tau)
=
a_\mu 
\Bigl(x_0^+  -i \sum_{i=1}^N G(\tau,\tau_i)k_i^+ \Bigr)
\, .
\label{abusesimp}
\ear
The master formula \eqref{Nphotonpw} can then be written more explicitly as
\bear
\Gamma_{\rm scal}(\lbrace{k_i,\varepsilon_i\rbrace};a)
\!\!\! &=& \!\!\! (-e)^N
(2\pi)^3 
\delta\bigl(\sum_{i=1}^N k_i^1\bigr)
\delta\bigl(\sum_{i=1}^N k_i^2\bigr)
\delta\bigl(\sum_{i=1}^N k_i^+\bigr)
\!\!
\totint dx_0^+ \e^{-i x_0^+ \sum_{i=1}^N k_i^-}
\nonumber\\
\hspace{-1em}&&\hspace{-50pt} \times
\int_0^\infty
\frac{dT}{T}\,
{(4\pi T)}^{-{D\over 2}}
\prod_{i=1}^N\int_0^Td\tau_i\,
\e^{-\bigl(m^2+ \langle\langle a^2 \rangle\rangle - \langle\langle a \rangle\rangle^2\bigr)T}
\nonumber\\
\hspace{-1em}&&\hspace{-50pt} \times
\mathfrak{P}_{N} 
\,{\rm e}^{\half \sum_{i,j=1}^N G_{ij}k_i\cdot k_j+2\sum_{i=1}^N k_i \cdot 
\bigl(I(\tau_i)-\langle\langle I \rangle\rangle \bigr)}
\label{GammaNexplpw}
\ear
where the polynomials $\mathfrak{P}_{N}$ are defined by  (compare \eqref{defPN})
\bear
{\rm e}^{\sum_{i,j=1}^N\big(-i\dot G_{ij}\varepsilon_i\cdot k_j+\half \ddot G_{ij}\varepsilon_i\cdot \varepsilon_j\big)
-2i \sum_{i=1}^N\bigl(a(\tau_i)-\langle\langle a\rangle\rangle \bigr) \cdot \varepsilon_i
}
\big \vert_{\varepsilon_{1}\cdots\varepsilon_{N}}
\equiv
(-i)^{N} {\mathfrak{P}}_{N} \, .
\label{defPNpw}
\ear
Proceeding to the spinor QED case, in the above-mentioned Fradkin approach \cite{fradkin}
one gets a representation of the $N$-photon amplitude with a fermion loop that differs from the
scalar-loop one \eqref{12-Nphotonpw} only by an additional Grassmann path integral,
\bear
-\half 
 \int D\psi \,
e^{-\int_0^T d\tau\bigl(
\half \psi\cdot \dot\psi
-ie \psi^{\mu}f_{\mn}\psi^{\nu}\bigr)}
\label{spinorQEDgrass}
\ear\no
where $ef_\mn = n_\mu a'_\nu - a'_\mu n_\nu$ is the field-strength tensor of the plane-wave field
and the path integral has to be taken with anti-periodic boundary conditions.

The photon vertex operator for the emission of a photon from a fermion line $V^{\gamma}_{\rm spin}$ differs from the scalar one 
\eqref{defphotonvertop} by a term representing the interaction of the fermion spin with
the photon, 
\bear
 V^{\gamma}_{\rm spin}[k,\varepsilon] &\equiv &  \int_0^Td\tau\, \Bigl[ \varepsilon\cdot \dot x(\tau)
 -i \psi(\tau)\cdot f[k,\varepsilon] \cdot \psi(\tau)\Bigr] \,\e^{ik\cdot x(\tau)}
\label{defVspin}
\ear
where $f[k,\varepsilon]\equiv k \otimes \varepsilon - \varepsilon \otimes k$. 

The same procedure that we applied above to eliminate $q(\tau)$ from the argument of $a_\mu$ works 
in this more general context to achieve the same change in the argument of $f_\mn$.
The Grassmann path-integral then becomes independent of the coordinate-space one. 
It is already gaussian and can be performed by Wick contractions with the basic correlator
\bear
\langle \psi^\mu(\tau) \psi^\nu(\tau') \rangle = \half \mathfrak G_F^\mn(\tau,\tau') 
\label{wickpsigen}
\ear
where the Green's function $ \mathfrak G_F$ is the inverse of the operator
\bear
{\cal O} \equiv  \frac{\delta_\mn}{2}  \ddtau + i a'_\mu(\tau)n_\nu  - i n_\mu a'_\nu (\tau) 
\label{defO}
\ear
with anti-periodic boundary conditions. With the further definitions
\bear
J_\mu(\tau) &\equiv& \int_0^\tau d\bar\tau \Bigl( a'_\mu(\bar\tau) - \langle\langle a'_\mu \rangle\rangle \Bigr)\, , \label{defJ}\\
{\cal J}_\mu(\tau,\tau') &\equiv& J_\mu(\tau)-J_\mu(\tau') - \frac{T}{2}\dot G (\tau,\tau')  \langle\langle a'_\mu \rangle\rangle 
\label{def calJ}
\ear
this Green's function can be compactly written as
\begin{equation}
\hspace{-1.5em}\mathfrak G_F^\mn(\tau,\tau') 
=
G_F(\tau,\tau')
\biggl\lbrace \delta^\mn + 2i n^\mu{\cal J}^\nu(\tau,\tau') + 2i {\cal J}^\mu(\tau',\tau)n^\nu
+ 2\Bigl\lbrack {\cal J}^2(\tau,\tau')-\frac{T^2}{4} \langle\langle a'\rangle\rangle^2\Bigr\rbrack  
n^\mu n^\nu\biggr\rbrace 
\, .
\label{Gfplane}
\end{equation}
The decomposition of the vertex operator \eqref{defVspin} into an orbital and a spin contribution
in the worldline formalism naturally leads to a spin-orbit decomposition of the $N$-photon amplitudes, 
\bear
\Gamma_{{\rm spin},N} = \sum_{S=0}^N \Gamma_{NS}\, ,\quad 
\Gamma_{NS} = \sum_{\lbrace i_1i_2\ldots i_S\rbrace} 
\Gamma_{NS}^{\lbrace i_1i_2\ldots i_S\rbrace}\, , 
\label{decompGammaSO}
\ear
where $S$ denotes the number of spin interactions, and the sum $ \sum_{\lbrace i_1i_2\ldots i_S\rbrace}$ 
runs over all choices of $S$ out of the $N$ photons as the ones assigned to those interactions. 
One arrives at the following master formula for $\Gamma_{NS}^{\lbrace i_1i_2\ldots i_S\rbrace}$ \cite{141}:
\bear
\Gamma_{NS}^{\lbrace i_1i_2\ldots i_S\rbrace}
&=& -2 (-e)^N
(2\pi)^3 
\delta\bigl(\sum_{i=1}^N k_i^1\bigr)
\delta\bigl(\sum_{i=1}^N k_i^2\bigr)
\delta\bigl(\sum_{i=1}^N k_i^+\bigr)
\totint dx_0^+ \e^{-i x_0^+ \sum_{i=1}^N k_i^-}
\nonumber\\&&\hspace{-50pt} \times
\int_0^\infty
\frac{dT}{T}\,
{(4\pi T)}^{-{D\over 2}}
\prod_{i=1}^N\int_0^Td\tau_i
\e^{-\bigl(m^2+ \langle\langle a^2 \rangle\rangle - \langle\langle a \rangle\rangle^2\bigr)T}
\,
\nonumber\\&&\hspace{-50pt} \times
\mathfrak{W} (k_{i_1},\varepsilon_{i_1};\ldots;k_{i_S},\varepsilon_{i_S}) 
\mathfrak{P}_{NS}^{\lbrace i_1i_2\ldots i_S\rbrace} 
{\rm e}^{\half \sum_{i,j=1}^N G_{ij}k_i\cdot k_j+2\sum_{i=1}^N k_i \cdot 
\bigl(I(\tau_i)-\langle\langle I \rangle\rangle \bigr)}
\, .
\label{GammaNSexplpw}
\ear
The polynomials $\mathfrak{P}_{NS}$ are defined by 
\bear
&&
{\rm e}^{\sum_{i,j=1}^N\big(-i\dot G_{ij}\varepsilon_i\cdot k_j+\half \ddot G_{ij}\varepsilon_i\cdot \varepsilon_j\big)
-2i \sum_{i=1}^N\bigl(a(\tau_i)-\langle\langle a\rangle\rangle \bigr) \cdot \varepsilon_i
}
\big \vert_{\varepsilon_{i_1}= \cdots = \varepsilon_{i_S} =0}\big \vert_{\varepsilon_{i_{S+1}}\cdots\varepsilon_{i_N}}
\nonumber\\
&& 
\equiv
(-i)^{N-S} {\mathfrak{P}}_{NS}^{\lbrace i_1i_2\ldots i_S\rbrace } \, ,
\label{defPNSpw}
\ear
where
\bear
\mathfrak{W} (k_{i_1},\varepsilon_{i_1};\ldots;k_{i_S},\varepsilon_{i_S})
\equiv  \Bigl\langle 
\psi_{i_1}\cdot f_{i_1} \cdot \psi_{i_1}
\cdots
\psi_{i_S}\cdot f_{i_S} \cdot \psi_{i_S}
\Bigr\rangle
\quad\quad
\label{defW}
\ear
and the notation on the left-hand side means that 
one first sets the polarisation vectors $\varepsilon_{i_1},\ldots,\varepsilon_{i_S}$ equal to zero,
and then selects all the terms linear in the surviving polarisation vectors. 

\section{$N$-photon amplitudes in a combined constant and plane-wave background (scalar QED)}

We return to the scalar QED case, and proceed to the combination of the plane-wave with a constant-field background.
Using Fock-Schwinger gauge, the constant field will add a contribution $\half ie\, x^{\mu} F_\mn \dot x^{\nu}$ to the
worldline Lagrangian, which can be absorbed into the modified Green's function \eqref{calGB} and the
field-dependent determinant factor shown in \eqref{masterF} \cite{shaisultanov,18}. We would now like to proceed as above 
and remove the path-integration variable from the argument of $a_\mu$. For this to go through,
we need to maintain the vanishing of the worldline correlators $\langle q(\tau)\cdot n \, q(\tau')\cdot n \rangle$ and
$\langle q(\tau) \cdot n\, \dot q(\tau') \cdot a(\tau')\rangle$, which requires a generalization of the identities $n^2=0$ and $n\cdot a=0$ to
\bear
n\cdot {\cal G}_B\cdot n = n\cdot \dot {\cal G}_B\cdot a =0 \, .
\label{idvan}
\ear
Unfortunately, it is easy to see that these identities cannot be true for a general combination of a plane-wave field and a constant one. 
It is for this reason that we now specialize our study to the case mentioned in the introduction, where the spatial part of $n$, 
the electric field, and the magnetic field are all parallel (this case will be simply called ``parallel case'' in the following).
This implies that, in a suitable Lorentz frame, $n$ becomes an eigenvector of $F_{\mu\nu}$,
$F_{\mu\nu}n^\nu \sim n_\mu$ \cite{frgish-book}, which further implies that 
\bear
n\cdot F^n \cdot n = n\cdot F^n \cdot a=0
\label{idvan2}
\ear
for any $n$ and, since
${\cal G}_B$ can be written as a power series in the field strength matrix, also \eqref{idvan}. Thus for this case one can proceed as 
in the pure plane-wave case, and arrive at the following generalization 
of the master formula \eqref{Nphotonpw},
\bear
\Gamma_{\rm scal}(\lbrace{k_i,\varepsilon_i\rbrace};a,F) &=&
(-ie)^N 
(2\pi)^3 
\delta\bigl(\sum_{i=1}^N k_i^1\bigr)
\delta\bigl(\sum_{i=1}^N k_i^2\bigr)
\delta\bigl(\sum_{i=1}^N k_i^+\bigr)
\totint dx_0^+ \e^{-i x_0^+ \sum_{i=1}^N k_i^-}
\nonumber\\&&\hspace{-80pt}\times
\int_0^{\infty}
\frac{dT}{T}\,
{(4\pi T)}^{-{D\over 2}}
{\rm det}^{{1\over 2}}
\biggl[
\frac{\cal Z}{{\rm sin}{\cal Z}}
\biggr]
\prod_{i=1}^N \int_0^Td\tau_i
\,{\rm e}^{\sum_{i,j=1}^N 
\bigl\lbrack \half k_i\cdot  {\cal G}_{Bij} \cdot k_j
-i\varepsilon_i \cdot \dot {\cal G}_{Bij}\cdot k_j
+\half\varepsilon_i\cdot \ddot {\cal G}_{Bij}\cdot\varepsilon_j
\bigr\rbrack}
\nonumber\\&&\hspace{-80pt}\times
\e^{-\bigl\lbrack m^2+ \frac{1}{2} \int_0^Td\tau\int_0^Td\tau' {\tilde a}(\tau)\cdot \ddot {\cal G}_B(\tau,\tau')\cdot {\tilde a}(\tau')\bigr\rbrack T
-\sum_{i=1}^N \int_0^Td\tau 
\bigl\lbrack {\tilde a}(\tau) \cdot\dot {\cal G}_B(\tau,\tau_i) \cdot k_i + i {\tilde a}(\tau) \cdot  \ddot {\cal G}_B (\tau,\tau_i) \cdot \varepsilon_i \bigr\rbrack
}
\Bigl\vert_{\varepsilon_1\cdots \varepsilon_N}
\nonumber\\
\label{NphotonpwF}
\ear
where now
\bear
{\tilde a}_\mu(\tau)
\equiv
a_\mu 
\Bigl(x_0^+  + n\cdot \sum_{i=1}^N\bigl [-i {\cal G}_B(\tau,\tau_i)\cdot k_i + \dot {\cal G}_B(\tau,\tau_i)\cdot \varepsilon_i\bigr]\Bigr)
\, .
\label{12-abusechanged}
\ear
Finally, we impose the gauge condition \eqref{polgauge}. In the parallel case this implies further that $n\cdot \dot {\cal G}_B \cdot \varepsilon_i =0$,
so that the polarizations disappear from the argument of $\tilde a_\mu$, and we arrive at a generalization of \eqref{GammaNexplpw},
\bear
\Gamma_{\rm scal}(\lbrace{k_i,\varepsilon_i\rbrace};a,F) &=&
(-ie)^N 
(2\pi)^3 
\delta\bigl(\sum_{i=1}^N k_i^1\bigr)
\delta\bigl(\sum_{i=1}^N k_i^2\bigr)
\delta\bigl(\sum_{i=1}^N k_i^+\bigr)
\totint dx_0^+ \e^{-i x_0^+ \sum_{i=1}^N k_i^-}
\nonumber\\&&\hspace{-80pt}
\times
\int_0^{\infty}
\frac{dT}{T}\,
{(4\pi T)}^{-{D\over 2}}
{\rm det}^{{1\over 2}}
\biggl[
\frac{\cal Z}{{\rm sin}{\cal Z}}
\biggr]
\prod_{i=1}^N \int_0^Td\tau_i
\,{\rm e}^{\half\sum_{i,j=1}^N k_i\cdot  {\cal G}_{Bij} \cdot k_j}
\nonumber\\&&\hspace{-80pt}\times
\e^{-\bigl(m^2+ \frac{1}{2} \int_0^Td\tau\int_0^Td\tau' {\tilde a}(\tau)\cdot \ddot {\cal G}_B(\tau,\tau')\cdot {\tilde a}(\tau'))T
-\sum_{i=1}^N \int_0^Td\tau 
{\tilde a}(\tau) \cdot\dot {\cal G}_B(\tau,\tau_i) \cdot k_i 
}
\,\widetilde{\mathfrak{P}}_{N}
\nonumber\\
\label{NphotonpwF}
\ear
with polynomials $\widetilde{\mathfrak{P}}_{N}$ defined by  
\bear
\,{\rm e}^{\sum_{i,j=1}^N 
\bigl\lbrack 
-i\varepsilon_i \cdot \dot {\cal G}_{Bij}\cdot k_j
+\half\varepsilon_i\cdot \ddot {\cal G}_{Bij}\cdot\varepsilon_j
\bigr\rbrack
-i\sum_{i=1}^N \int_0^Td\tau 
{\tilde a}(\tau) \cdot  \ddot {\cal G}_B (\tau,\tau_i) \cdot \varepsilon_i }
\Big \vert_{\varepsilon_{1}\cdots\varepsilon_{N}}
\equiv
(-i)^{N} {\widetilde{\mathfrak{P}}}_{N} \, .
\label{defPNpw}
\ear

\section{$N$-photon amplitudes in a combined constant and plane-wave background (spinor QED)}

Proceeding to spinor QED, we need to work out the changes caused by the constant field for the Grassmann path integral.
It adds a term $-ie\psi^\mu F_\mn \psi^\nu$ to the worldline Lagrangian which changes the Grassmann path integral normalization by
the determinant factor ${\rm det}^\half [{\cos}{\cal Z}]$, familiar from the Euler-Heisenberg Lagrangian, and 
generalizes the kinetic operator \eqref{defO} to
\bear
\widetilde {\cal O} \equiv  \frac{\delta_\mn}{2}  \ddtau + i {\tilde a}'_\mu(\tau)n_\nu  - i n_\mu {\tilde a}'_\nu (\tau) -ieF_\mn \, .
\label{deftildeO}
\ear
Thus we need the corresponding generalization of the Green's function \eqref{Gfplane}. 
Denoting this function by $\widetilde{\mathfrak{G}}_F$, we construct it by the expansion
\bear
\widetilde {\cal O}^{-1} =  \Bigl(\frac{1}{2}  \ddtau  -ieF\Bigr)^{-1} +
\Bigl(\frac{1}{2}  \ddtau  -ieF\Bigr)^{-1}
i\Bigl( n \otimes {\tilde a}' -  {\tilde a}' \otimes n \Bigr)
\Bigl(\frac{1}{2}  \ddtau  -ieF\Bigr)^{-1}
+ \ldots
\nonumber\\
\ear
and noting that only the first three terms of this series are non-vanishing due to the relations \eqref{idvan2}.
This leads to 
\bear
\widetilde{\mathfrak{G}}_F(\tau,\tau') &=& {\cal G}_F(\tau,\tau') 
\nonumber\\
&& 
+ i\int_0^Td\tau_1 
\Bigl\lbrack {\cal G}_F(\tau,\tau_1)\cdot n\, {\tilde a}'(\tau_1)\cdot {\cal G}_F(\tau_1,\tau') 
- {\cal G}_F(\tau,\tau_1) \cdot {\tilde a}'(\tau_1) n\cdot {\cal G}_F (\tau_1,\tau') \Bigr\rbrack
\nonumber\\
&& + \int_0^Td\tau_1\int_0^Td\tau_2 \, {\cal G}_F(\tau,\tau_1)\cdot n \, {\tilde a}'(\tau_1) \cdot {\cal G}_F(\tau_1,\tau_2)
\cdot {\tilde a}'(\tau_2) n\cdot {\cal G}_F(\tau_2,\tau')
\, .
\nonumber\\
\label{tildeGF}
\ear
Note the anti-symmetry relation $\widetilde {\mathfrak{G}}_F(\tau',\tau)  = -\widetilde {\mathfrak{G}}_F^{\intercal}(\tau,\tau')$. 
With the Green's function in hand, it is immediate to generalize the master formula \eqref{GammaNSexplpw} to 
\bear
\Gamma_{NS}^{\lbrace i_1i_2\ldots i_S\rbrace}
&=& -2 (-e)^N
(2\pi)^3 
\delta\bigl(\sum_{i=1}^N k_i^1\bigr)
\delta\bigl(\sum_{i=1}^N k_i^2\bigr)
\delta\bigl(\sum_{i=1}^N k_i^+\bigr)
\totint dx_0^+ \e^{-i x_0^+ \sum_{i=1}^N k_i^-}
\nonumber\\&&\hspace{-50pt}
\times
\int_0^{\infty}
\frac{dT}{T}\,
{(4\pi T)}^{-{D\over 2}}
{\rm det}^{{1\over 2}}
\biggl[
\frac{\cal Z}{{\rm tan}{\cal Z}}
\biggr]
\prod_{i=1}^N \int_0^Td\tau_i
\,{\rm e}^{\half\sum_{i,j=1}^N k_i\cdot  {\cal G}_{Bij} \cdot k_j}
\nonumber\\&&\hspace{-50pt}\times
\e^{-\bigl\lbrack m^2+ \frac{1}{2} \int_0^Td\tau\int_0^Td\tau' {\tilde a}(\tau)\cdot \ddot {\cal G}_B(\tau,\tau')\cdot {\tilde a}(\tau')\bigr\rbrack T
-\sum_{i=1}^N \int_0^Td\tau 
{\tilde a}(\tau) \cdot\dot {\cal G}_B(\tau,\tau_i) \cdot k_i 
}
\nonumber\\&&\hspace{-50pt} \times
\widetilde{\mathfrak{W}} (k_{i_1},\varepsilon_{i_1};\ldots;k_{i_S},\varepsilon_{i_S}) 
\widetilde{\mathfrak{P}}_{NS}^{\lbrace i_1i_2\ldots i_S\rbrace} 
\, .
\label{GammaNSexplpwF}
\ear
The polynomials $\widetilde{\mathfrak{P}}_{NS}$ are now defined by 
\bear
&&
\,{\rm e}^{\sum_{i,j=1}^N 
\bigl\lbrack 
-i\varepsilon_i \cdot \dot {\cal G}_{Bij}\cdot k_j
+\half\varepsilon_i\cdot \ddot {\cal G}_{Bij}\cdot\varepsilon_j
\bigr\rbrack
-i\sum_{i=1}^N \int_0^Td\tau 
{\tilde a}(\tau) \cdot  \ddot {\cal G}_B (\tau,\tau_i) \cdot \varepsilon_i }
\Big \vert_{\varepsilon_{1}\cdots\varepsilon_{N}}
\big \vert_{\varepsilon_{i_1}= \cdots = \varepsilon_{i_S} =0}\big \vert_{\varepsilon_{i_{S+1}}\cdots\varepsilon_{i_N}}
\nonumber\\
&& 
\equiv
(-i)^{N-S} {\widetilde{\mathfrak{P}}}_{NS}^{\lbrace i_1i_2\ldots i_S\rbrace } \, ,
\label{defPNSpw}
\ear
and the calculation of $\widetilde{\mathfrak{W}}$ proceeds like the one of $\mathfrak{W}$ in section \ref{pw},
only that now the Green's function $\widetilde{\mathfrak{G}}_F$ must be used in \eqref{defW}.

Note that for $N=0$ one has ${\tilde a}_{\mu} = a_\mu (x_0)$, which leads to the vanishing of the integral
$ \int_0^Td\tau\int_0^Td\tau' {\tilde a}(\tau)\cdot \ddot {\cal G}_B(\tau,\tau')\cdot {\tilde a}(\tau') $. 
Thus in the photonless case \eqref{GammaNSexplpwF} reduces to \eqref{masterF}, i.e. to the Euler-Heisenberg action,
without any effect of the plane-wave field \cite{schwinger51}.

\section{Explicit representation of the generalized worldline Green's functions}

It is known from the constant-field case that using the master formula \eqref{masterF} for actual calculations in most cases requires
a more practical representation of the generalized worldline Green's functions ${\cal G}_{B,F}$ than the definition as power series 
\eqref{calGB}, \eqref{calGF}. Suitable formulas have been derived in a Lorentz-invariant way \cite{shaisultanov,40}, but here we will make use of the Lorentz frame 
that is part of the definition of the ``parallel'' special case of the constant plus plane-wave field, and assume that ${\bf n},{\bf E}, {\bf B}$ 
are all pointing along the positive $z$ - axis. In our euclidean conventions we then have $F = Br_{\perp} + iEr_{\parallel}$, where
\begin{equation}
r_{\perp} \equiv
\left(
\begin{array}{*{4}{c}}
0&1&0&0\\
-1&0&0&0\\
0&0&0&0\\
0&0&0&0
\end{array}
\right),\qquad
r_{\parallel} \equiv
\left(
\begin{array}{*{4}{c}}
0&0&0&0\\
0&0&0&0\\
0&0&0&1\\
0&0&-1&0
\end{array}
\right)
\, .
\nonumber\\
\label{defr}\nonumber
\non\\
\vspace{5mm}
\end{equation}
and the eigenvalue equation 
\bear
F\cdot n = -E n
\, .
\label{evn}
\ear

We will also need projection matrices $g_{\perp},g_{\parallel}$, defined by
\begin{equation}
g_{\perp} \equiv
\left(
\begin{array}{*{4}{c}}
1&0&0&0\\
0&1&0&0\\
0&0&0&0\\
0&0&0&0
\end{array}
\right),\qquad
g_{\parallel} \equiv
\left(
\begin{array}{*{4}{c}}
0&0&0&0\\
0&0&0&0\\
0&0&1&0\\
0&0&0&1
\end{array}
\right)
\, .
\nonumber\\
\label{defgs}\nonumber
\vspace{7 mm}
\end{equation}
Following \cite{40}, we can decompose the power series \eqref{calGB}, \eqref{calGF} into their even (odd) parts ${\cal S}_{B,F}$ (${\cal A}_{B,F}$),
$
{\cal G}_{B,F}
=
{\cal S}_{B,F}
+
{\cal A}_{B,F}
$,
and those further into
\bear
{\cal S}_{B12}^{\mu\nu}
&=&
-{T\over 2}
\sum_{\alpha ={\perp},{\parallel}}
{A_{B12}^{\alpha}\over z_{\alpha}}\,g_{\alpha}^{\mu\nu}
\nonumber\\
{\cal A}_{B12}^{\mu\nu}
&=&
{iT\over 2}
\sum_{\alpha ={\perp},{\parallel}}
{S_{B12}^{\alpha}-\dot G_{B12}\over z_{\alpha}}
\,r_{\alpha}^{\mu\nu}
\nonumber\\
{\cal S}_{F12}^{\mu\nu} &=&
\sum_{\alpha ={\perp},{\parallel}}
S_{F12}^{\alpha}\,g_{\alpha}^{\mu\nu}
\nonumber\\
{\cal A}_{F12}^{\mu\nu} &=& 
-i
\sum_{\alpha ={\perp},{\parallel}}
A_{F12}^{\alpha}\,r_{\alpha}^{\mu\nu}
\nonumber\\
\label{specialdecompcalSA}
\ear\no
with $z_{\perp}=eBT$, $z_{\parallel} = ieET$ and the 
scalar, dimensionless coefficient functions
\bear
S_{B12}^{\alpha} &=&
{\sinh(z_{\alpha}\dot G_{B12})\over \sinh(z_{\alpha})} 
\, ,\quad
A_{B12}^{\alpha} =
{\cosh(z_{\alpha} \dot G_{B12})\over 
\sinh(z_{\alpha})}-{1\over z_{\alpha}}
\nonumber\\
S_{F12}^{\alpha} &=&
G_{F12}{\cosh(z_{\alpha}\dot G_{B12})\over\cosh(z_{\alpha})}
\, , \quad 
A_{F12}^{\alpha} =
G_{F12}{\sinh(z_{\alpha}\dot G_{B12})\over \cosh(z_{\alpha})}
\nonumber\\
\label{defAB}
\ear\no
%
Before working out the right-hand side of \eqref{tildeGF}, let us invoke gauge invariance once more to make the plane-wave field
fully transversal. This allows us to write
\bear
{\tilde a}'={\tilde a}'_+ m_+ + {\tilde a}'_- m_-
\ear
where $m_\pm \equiv \frac{1}{\sqrt{2}} (1,\pm i,0,0)$ are a basis of transversal vectors that are also eigenvectors of $F$,
\bear
F\cdot m_\pm = \pm iB m_\pm
\label{Fm}
\ear
Further, it will be convenient to introduce (compare \eqref{defJ})
\bear
\tilde J^{\mu}(\tau) &\equiv& \sum_{\lambda =\pm}m_{\lambda}^\mu
 \e^{-2\frac{\tau}{T} (z_{\parallel} + \lambda z_{\perp})}
 \int_0^\tau d\bar\tau
 \Bigl( {\tilde a}'_{\lambda} (\bar\tau) - \langle\langle  {\tilde a}'_{\lambda} \rangle\rangle_F \Bigr)
 \e^{2\frac{\bar\tau}{T} (z_{\parallel} + \lambda z_{\perp})}
 \nonumber\\
 \label{defJtilde}
\ear
where 
\bear
 \langle\langle  {\tilde a}'_{\lambda} \rangle\rangle_F \equiv 
 \frac{\int_0^Td\tau {\tilde a}'_{\lambda}(\tau) \e^{-2 (z_{\parallel} + \lambda z_{\perp}) \frac{T-\tau}{T}}}
 {\int_0^T d\tau \e^{-2 (z_{\parallel} + \lambda z_{\perp}) \frac{T-\tau}{T}}}
 =
\frac{2 (z_{\parallel} + \lambda z_{\perp})}{1-e^{-2 (z_{\parallel} + \lambda z_{\perp})}}
 \frac{1}{T}
  \int_0^Td\tau {\tilde a}'_{\lambda}(\tau) \e^{-2 (z_{\parallel} + \lambda z_{\perp}) \frac{T-\tau}{T}}
  \nonumber\\
 \label{defFav}
 \ear
We note the following properties of $\tilde J^\mu (\tau)$:

\begin{enumerate}

\item

$
\tilde J^\mu (0) = \tilde J^\mu (T) = 0
$

\item
$\tilde J^\mu(\tau)$ turns into $J^\mu(\tau)$ for $F=0$

\item
\bear
n\otimes \Bigl( {\tilde a}'(\tau) - \langle\langle {\tilde a}'\rangle\rangle_F  \Bigr)
&=&
n\otimes \dot {\tilde J} (\tau) -(2iF)\cdot n \otimes  {\tilde J} (\tau) + n\otimes \tilde J(\tau) \cdot (2iF)
\nonumber\\
&=&
\Bigl( \ddtau  - \bigl\lbrack 2iF,\cdot \bigr\rbrack \Bigr) n\otimes {\tilde J}(\tau)
\label{pretty}
\ear

\end{enumerate}

We apply this identity to all the factors of $n\otimes {\tilde a}'$ appearing in \eqref{tildeGF}, and then remove all derivatives 
of $\tilde J$ by suitable integrations-by-parts. After a multiple application of the defining equation of the generalized 
worldline Green's function ${\cal G}_F$,
$( \ddtau  - 2iF) {\cal G}_F (\tau,\tau') = 2\delta(\tau-\tau')\Eins$, to remove some integrations,
the remaining ones can be done using the remarkable integral formula
\bear
&&
\int_0^Td\tau_1 G_F(\tau,\tau_1) 
\frac{\e^{z_\parallel \dot G(\tau,\tau_1)}} {\cosh z_\parallel}
G_F(\tau_1,\tau') 
\frac{\e^{-\lambda z_{\perp} \dot G(\tau_1,\tau')}}{\cosh z_\perp }
= \nonumber\\
 &&-\frac{T}{z_\parallel + \lambda z_\perp} G_F(\tau,\tau')
\biggl\lbrack
\frac{\e^{z_\parallel \dot G(\tau,\tau')}}{\cosh z_\parallel} 
-
\frac{\e^{ -\lambda z_\perp \dot G(\tau,\tau')}}{\cosh z_\perp} 
\biggr\rbrack
\ear
The right-hand side of \eqref{tildeGF} then appears written in terms of $\tilde J^\mu$ and ${\cal G}_F$ as follows, 
\bear
\widetilde{\mathfrak{G}}_F(\tau,\tau') &=& 
{\cal G}_F (\tau,\tau') 
+ 2i
\Bigl\lbrack
n\otimes \tilde J(\tau) \cdot {\cal G}_F(\tau,\tau')
-{\cal G}_F(\tau,\tau')\cdot n \otimes
{\tilde J} (\tau')
\Bigr\rbrack
\nonumber\\
&&. \hspace{-30pt}
+2i
\Bigl\lbrack
{\cal G}_F(\tau,\tau') \cdot {\tilde J}(\tau') \otimes n
- {\tilde J} (\tau) \otimes n\cdot {\cal G}_F (\tau,\tau')
\Bigr\rbrack
\nonumber\\
&& \hspace{-30pt}
+ 2 {\tilde J}^2(\tau) n \otimes n\cdot {\cal G}_F(\tau,\tau') 
+ 2 {\cal G}_F(\tau,\tau')\cdot n \otimes n {\tilde J}^2(\tau')
\nonumber\\
&&\hspace{-30pt}
- 4 {\tilde J}(\tau) \cdot {\cal G}_F(\tau,\tau') \cdot {\tilde J}(\tau') n\otimes n
\nonumber\\
&&\hspace{-30pt}
-\frac{iT}{z_\parallel + \lambda z_\perp}
\Bigl\lbrack
{\cal G}_F(\tau,\tau')\cdot
\Bigl( n\otimes \langle\langle{\tilde a_\lambda}'\rangle\rangle_F m_\lambda -  \langle\langle{\tilde a_{\lambda}}'\rangle\rangle_F m_\lambda \otimes n \Bigr)
\nonumber\\
&& \hspace{60pt}
-
\Bigl( n\otimes \langle\langle{\tilde a_\lambda}'\rangle\rangle_F m_\lambda -  \langle\langle{\tilde a_\lambda}'\rangle\rangle_F m_\lambda \otimes n \Bigr)
\cdot {\cal G}_F(\tau,\tau')
\Bigr\rbrack
\nonumber\\
&&\hspace{-30pt}
 + 2\frac{T}{z_\parallel + \lambda z_\perp}\Bigl\lbrack
\langle\langle{\tilde a_\lambda}'\rangle\rangle_Fm_\lambda \cdot \tilde J(\tau'){\cal G}_F(\tau,\tau') \cdot n \otimes n 
 + \langle\langle{\tilde a_\lambda}'\rangle\rangle_Fm_\lambda \cdot \tilde J(\tau) n \otimes n \cdot {\cal G}_F(\tau,\tau')\nonumber\\
&& - (\tilde J (\tau) \cdot {\cal G}_F(\tau,\tau') \cdot m_\lambda\langle\langle{\tilde a_\lambda}'\rangle\rangle_F 
+ \langle\langle{\tilde a_\lambda}'\rangle\rangle_F m_\lambda \cdot  {\cal G}_F(\tau,\tau') \cdot \tilde J(\tau') )
n\otimes n
\Bigr\rbrack
 \nonumber\\
&&\hspace{-30pt}
+ \frac{T^2}{z_{\parallel}^2-z_{\perp}^2}
\Bigl\lbrack 
\frac{ \langle\langle{\tilde a}'\rangle\rangle_F^2}{2} 
{\cal G}_F(\tau,\tau')\cdot  n\otimes n 
+ \frac{ \langle\langle{\tilde a}'\rangle\rangle_F^2}{2} 
 n\otimes n \cdot {\cal G}_F(\tau, \tau') 
 \nonumber\\&& \hspace{40pt}
- \langle\langle{\tilde a}'\rangle\rangle_F \cdot {\cal G}_F(\tau ,\tau') \cdot \langle\langle{\tilde a}'\rangle\rangle_F
n\otimes n
\Bigr\rbrack
\nonumber\\
\ear
Here it is understood that the index $\lambda$ is always summed over. 
%
%
This way of writing the Green's function makes it also easy to see that it fulfills the basic properties of anti-symmetry and anti-periodicity, and that it identifies with
the right-hand side of \eqref{Gfplane} in the limit $F_\mn \to 0$.
We note that some simplification is possible for $E=0$ since then ${\cal G}_F(\tau,\tau')\cdot n =  G_F(\tau,\tau') n$.

\section{Summary and Outlook}

We have used the worldline formalism to derive master formulas for the $N$-photon amplitudes in a background field
that is a combination of a constant field with a plane-wave one, for the special ``parallel'' configuration that has been known
to be particularly amenable to an analytical treatment. As usual in QED applications, the formalism has allowed us to unify the scalar and spinor QED cases, 
and to significantly reduce the algebraic complexity by avoiding the break-up of the scalar or fermion loop into individual propagators. 
All the effects of the addition of the constant external field have been absorbed into generalizations of the various worldline Green's functions,
global determinant factors and the single quantity ${\tilde J}^\mu (\tau)$ defined in \eqref{defJtilde}. 
These advantages over the standard diagrammatic approach will naturally become increasingly more pronounced with increasing number of photons. 

In a separate publication we will present the application of the master formula to the $N=2$ case, that is, the vacuum polarization
and associated helicity flip in scalar and spinor QED.

Further generalization from the closed-loop to the open-line case should be feasible along the lines of \cite{110,130,131}.

\no
{\bf Acknowledgements:} This work was supported by the project Advanced Research using High Intensity 
Laser Produced Photons and Particles (ADONIS) (Project No. $CZ.02.1.01/0.0/0.0/16\_019/0000789$) from the European Regional Development Fund.

%



\end{document}